\documentclass[12pt]{article}

\advance\voffset by -2.0cm \advance\hoffset by -1.25cm
\usepackage{amsmath}
\usepackage{amssymb}
\usepackage{amsthm}
\usepackage{amsfonts}

\numberwithin{equation}{section}

\theoremstyle{definition}

\newcommand{\QQ}{\mathbb{Q}} 
\newcommand{\ZZ}{\mathbb{Z}} 
\newcommand{\PP}{\mathbb{P}} 

\DeclareMathOperator{\Sym}{Sym} 

\newcommand{\I}{\mathcal{I}}
\newcommand{\Z}{\mathcal{Z}}
\newcommand{\F}{\mathcal{F}} 
\newcommand{\N}{\mathcal{N}} 
\newcommand{\J}{\mathcal{J}} 

\newcommand{\M}{{\mathcal M}} 
\newcommand{\be}{\begin{equation}}
\newcommand{\ee}{\end{equation}}

\def\1{\frak 1}
\def\2{\frak 2}
\def\3{\frak 3}

\newlength{\oldcolsep}

\begin{document}

\title{Finite Strings From Non-Chiral Mumford Forms}

\author{Marco Matone}\date{}

\maketitle

\begin{center} Dipartimento di Fisica ``G. Galilei'' and Istituto
Nazionale di Fisica Nucleare \\
Universit\`a di Padova, Via Marzolo, 8 -- 35131 Padova,
Italy\end{center}

\bigskip

\begin{abstract}
\noindent
We show that there is an infinite class of partition functions with world-sheet metric, space-time coordinates and first order systems, that correspond to volume forms on the moduli space
of Riemann surfaces and are free of singularities
at the Deligne-Mumford boundary. An example is the partition function with $4=2(c_2+c_3+c_4-c_5)$ space-time coordinates, a $b$-$c$ system of weight 3, one of weight 4 and a $\beta$-$\gamma$ system of weight 5.
Such partition functions are derived from the mapping of the Mumford forms to non-factorized scalar forms on $\M_g$ introduced in arXiv:1209.6049.
\end{abstract}

\newpage

\section{Introduction}

Let us start by considering the following suggestion. Instead of considering the problems of infinities on
the moduli space of curves $\M_g$ of a given string theory, we consider the opposite question. Essentially, we consider the
problem of finding the partition functions with a world sheet metric, $D$ scalars, $b$-$c$ and $\beta$-$\gamma$ systems
corresponding to volume forms on the moduli space of Riemann surfaces which are free of singularities
at the Deligne-Mumford boundary. Such a question has a positive answer in view of the results of \cite{Matone:2012kk}, where it was shown how to absorb, in a modular invariant way, the point dependence
due to the insertion of $b$ zero modes. As a consequence of that, the resulting partition functions involving the path integration on the world-sheet metric are well-defined volume forms $\M_g$.

The finding in \cite{Matone:2012kk}\cite{Matone:2012wy} are in some sense a consequence of the basic observation by Yu. I. Manin \cite{Manin:1986gx}\cite{BeilinsonZW}
that the bosonic Polyakov measure \cite{Polyakov:1981rd} is (essentially) the same of the modulo square of the
Mumford form \cite{Mumford}.

The formulation of string and superstring perturbation theories face the hard problem of considering their effective
characterization as geometrical objects on $\M_g$. The algebraic-geometrical structure of such a space, and of its Deligne-Mumford compactification $\overline\M_g$, provides a powerful
tool, still largely unexplored, to check the consistency of the theory at any genus. Several questions concerning the definition of the theory, such as unitarity and structure of the singularities, are endowed in the geometry
of $\overline\M_g$. Generally it is not true the opposite, there are subtle questions concerning $\overline\M_g$ which may not appear in the purely field theoretic formulation.
In this respect, in recent years it has been clear that the original formulations of the known string theories may map
to other formulations whose elementary fields are different. The main example is the Berkovits pure spinor formulation of superstring theory \cite{BerkovitsFE}\cite{geometrica}. Changing the field content leaving invariant
the theory is a question strictly related to the Mumford isomorphism
\be
\lambda_n\cong\lambda_1^{\otimes c_n}\ ,
\label{eccoliso}\ee
where $\lambda_n$ are the determinant line bundles and
$c_n=6n^2-6n+1$.
Physically $\lambda_n$ is associated to the partition function of a $b$-$c$ system of weight $n$, whereas the scalar target coordinates correspond
to $\lambda_1$, the Hodge bundle. For each $n$ the Mumford form $\mu_{g,n}$  is the unique, up to a constant, holomorphic section of $\lambda_n\otimes\lambda_1^{-c_n}$ nowhere vanishing on ${\cal M}_g$.
It is useful to note that this implies that its inverse is also holomorphic.

\noindent
Let $C$ be a compact Riemann surface of genus $g\geq2$. A further development concerning the Mumford isomorphism is the following map
\be
\psi:\Sym^n H^0(K_C)\longrightarrow H^0(K_C^n) \ ,
\label{mapsymm}\ee
which is surjective for $g=2$
and for $C$ non-hyperelliptic of genus $g>2$. Roughly speaking this provides another relation between the two sides of the Mumford isomorphism (\ref{eccoliso}).
It can be seen as a way to use the stringent properties of the Mumford forms $\mu_{g,n}$ in considering the problem of characterizing the Jacobian (\ref{mapsymm}).
This is the essence of recent developments in the characterization of Riemann surfaces in $\PP^{g-1}$ and the related Schottky problem
\cite{Matone:2005bx}\cite{Matone:2006bb}\cite{Matone:1900zz}\cite{Matone:2011ic}. It has been shown in \cite{Matone:2012wy}\cite{Matone:2012kk}
that the map (\ref{mapsymm}), together with the Mumford isomorphism, is crucial in constructing the string measures.

Here we show that the partition functions introduced in \cite{Matone:2012kk} include a class which is free of the singularities that usually arise when one considers
the pinching of handles of the Riemann surface.

\section{Singularity-free at the Deligne-Mumford boundary}

In \cite{Matone:2012kk} it has been considering string
theories with some underlying hidden symmetry which are still to be discovered from first principles. The preliminary basic step
has been to classify all forms on $\M_g$ satisfying some natural properties.

\begin{enumerate}

\item Since each string theory involves the path integration over
the world-sheet metric, it should be a modular invariant $(3g-3,3g-3)$ form, i.e. a volume form on $\M_g$.

\item Such forms should correspond to determinants of laplacians
associated to the space-time coordinates to $b$-$c$ and/or $\beta$-$\gamma$ systems of any conformal weight.

\item The combination of such determinants should be Weyl invariant.

\end{enumerate}

\noindent The solution is based on the Weyl and modular invariant properties of the Bergman reproducing kernel
\be
B(z,\bar w)=\sum_{1}^g\omega_j(z)(\tau_2^{-1})_{jk}\bar\omega_k(w)  \ ,
\label{brk}\ee
where $\omega_1,\ldots,\omega_g$, are the canonically normalized holomorphic abelian differentials, $\tau_2\equiv {\rm Im}\tau$,
with $\tau$ the Riemann period matrix.

Let us show that
\be
\int_{\M_g}{\cal Z}[\J]=\int DgDXD\Psi \exp(-S[X]-S[\Psi]) \ ,
\label{formidabile}\ee
corresponds to a volume form on $\M_g$. Here $S[X]$ denotes the Polyakov action in
$$D=26+2\sum_{k\in \I}n_kc_k \ ,$$
dimensions, where $c_k=6k^2-6k+1$ is (minus) 1/2 the central charge
of the non-chiral ($b$-$c$) $\beta$-$\gamma$ system of weight $k$.
$\I$ is the set of conformal weights $k\in\QQ$, $\J$ the set of
$n_k\in\ZZ/2$. $D\Psi$ denotes the product on $k\in \I$ of $|n_k|$
copies of the non-chiral measures, including the zero mode
insertions, of weight $k$ $b$-$c$ systems for $n_k>0$, or
$\beta$-$\gamma$ systems for $n_k<0$. $S[\Psi]$ is the sum of the
corresponding non-chiral $b$-$c$ and $\beta$-$\gamma$ actions.

The string partition functions ${\cal Z}[\J]$
correspond to the Polyakov partition function $\Z_{Pol}$ times a
rational function of determinants of laplacians. Namely \be
\Z[\J]=\Z_{Pol}\prod_{k\in \I} \Z_k^{n_k} \ .
\label{laprimaaaAAA}\ee
In particular, it turns out that \cite{Matone:2012kk} \be
\Z_n=\int DX Db D\bar b Dc D\bar c(b\bar b)_n\exp(-S[X]
-{1\over2\pi}\int_C\sqrt g b\nabla^z_{1-n}c+c.c.) \ ,
\label{nove}\ee with
$$
(b\bar b)_n= \int_{C^{N_n}} \prod_j B^{1-n}(z_j,\bar z_j)b(z_j)\bar
b(z_j) \ ,
$$
coincides with \be \Z_n=\Bigg({{\det}'\Delta_{0}\over{\cal
N}_0}\Bigg)^{-c_n} {\det \M_n\det '\Delta_{1-n}\over\det{\cal N}_n}
\ , \label{questionnnsolvedYYY}\ee
where
$$
(\M_n)_{jk}=\int_C \bar \phi_j^n(z) B^{1-n}(z,\bar z) \phi_k^n(z) \ ,
$$
$$
(\N_n)_{jk}=\int_C \bar\phi_j^n \rho^{1-n} \phi_k^n \ ,
$$
with $\phi^n_j$
are the zero modes associated to the $b$-field and
 $\rho=2g_{z\bar z}$ is the metric tensor in local complex coordinates, that is $ds^2=2g_{z\bar z}dz d \bar z$.
It follows by
(\ref{laprimaaaAAA}) that the string partition functions
(\ref{formidabile}) correspond to integral of Weyl invariant volume forms on
$\M_g$ \cite{Matone:2012kk}. This provides a manifestly intrinsic way to absorb the point
dependence, that is the expression
$$
B={}^t\phi^1\cdot Y^{-1}\cdot
\bar\phi^1 \ ,
$$ implies that $(b\bar b)_n$
 depends only on the complex
structure of the Riemann surface $C$.

Another result in \cite{Matone:2012kk} is that $\Z[\J]$ can be expressed in terms of the building blocks of the Mumford forms $F_{g,k}$. In particular,
\be \Z[\J]={|F_{g,2}|^2\over
(\det\tau_2)^{13}}\prod_{k\in\I} \Bigg({|F_{g,k}|^2\det\M_k\over
(\det\tau_2)^{c_k}}\Bigg)^{n_k}|\wedge^{3g-3}\phi_j^2|^2 \ ,
\label{dfyhg}\ee which also
provides the expression of $\Z[\J]$ in terms of theta functions.

Let us shortly illustrate the Mumford forms.
Let $\phi_{1}^n,\ldots,\phi_{N_n}^n$ be a basis of $H^0(K_C^n)$,
$$N_n=(2n-1)(g-1)+\delta_{n1} \ .$$
Following Fay, we set \cite{Fay}
$$
\kappa[\phi^1]={\det\phi_i^1(z_j)\sigma(y)\prod_1^gE(y,z_i)\over
\theta\bigl(\sum_{1}^gp_i-y-\Delta\bigr)\prod_1^g\sigma(z_i)
\prod_{i<j}^gE(z_i,z_j) } \ ,
$$
and, for $n>1$
$$
\kappa[\phi^n]={\det \phi_i^{n}(z_j)\over
\theta[\delta]\bigl(\sum_{1}^{N_n}
z_i-(2n-1)\Delta\bigr)\prod_{1}^{N_n}\sigma(p_i)^{2n-1}\prod_{i<j}^{N_n}
E(z_i,z_j)}  \ ,
$$
where the above notation about theta functions, prime form and the $g/2$-differential $\sigma$ is the standard one. Note that we added a dependence on the theta characteristic. Similar expressions hold for
more general cases such as non-integer $n$ \cite{FayMAM}. Consider the universal curve $\mathcal{C}_g$, that is
the $(3g-2)$-dimensional complex space built by placing over each point of $\M_g$ the corresponding curve $C$.
Let $\pi$ be the map projecting $\mathcal{C}_g$ to $\M_g$, and  $L_n=R\pi_*(K^n_{\mathcal{C}_g/\mathcal{M}_g})$ the vector bundle on $\mathcal{M}_g$ of rank
$N_n$ with fiber $H^0(K_C^n)$ at the point of $\mathcal{M}_g$ representing $C$.
Denote by $\lambda_n=\det L_n$ the determinant line bundle.
It turns out that, for each $n$, the form associated to the Mumford isomorphism is
\be
\mu_{g,n}=F_{g,n}[\phi^n]{\phi^n_1\wedge
\cdots\wedge\phi_{N_n}^n\over(\omega_1\wedge\cdots\wedge\omega_g)^{c_n}}
\ , \label{Mumfordformsss}\ee where \be
F_{g,n}[\phi^n]={\kappa[\omega]^{(2n-1)^2}\over \kappa[\phi^n]} \ .
\label{star}\ee

Such quantities naturally appear in describing the geometry of curves in $\PP^{g-1}$ and in characterizing the Jacobian locus \cite{Matone:2006bb}\cite{Matone:1900zz}\cite{ShepherdBarronHC}\cite{Matone:2011ic}.
These also appear in the string measures.
In \cite{Manin:1986gx}\cite{BeilinsonZW}\cite{Belavin:1986cy}\cite{FayMAM} it has been shown that
\be
\Bigg({{\det}'\Delta_{0}\over{\cal N}_0\det{\cal N}_1}\Bigg)^{-c_n}{{\det}'\Delta_{1-n}\over\det {\cal N}_{1-n}\det {\cal N}_n}=
\Bigg|{\kappa[\omega]^{(2n-1)^2}\over
\kappa[\phi^n]}\Bigg|^2 \ .
\label{bosstrinmeasBBB}\ee
Eq.(\ref{bosstrinmeasBBB}) expresses determinants of laplacians in terms of theta functions whereas Eq.(\ref{Mumfordformsss}) establishes the relation between determinants of laplacians and the Mumford forms. Let us note that there is a lot of literature on string determinants, loop amplitudes and $b$-$c$ systems, see for example \cite{D'Hoker:1986zt}-\cite{Bonora:1989zm}. In the following we will consider $n$ such that $h^0(K_C^{1-n})=0$ and drop
the term $\det {\cal N}_{1-n}$ in (\ref{bosstrinmeasBBB}).

The above prescription of absorbing the zero modes is essentially the only well-defined recipe on any Riemann surface. Nevertheless, there is a related approach which is defined on
the compact Riemann surfaces of genus two and on the non-hyperelliptic compact Riemann surfaces with $g>2$, these are called canonical curves.
Instead of integrating with $B^{1-n}(z_j,\bar z_j)$ each pair $b(z_j)\bar b(z_j)$ of the zero mode insertions, one may divide
them by the determinant of $B^{(n)}(z_j,\bar z_k)$, denoting the $n$-fold Hadamard product of
$B(z_j,\bar z_k)$. In particular, it turns out that
 \be V_n=\int DX Db D\bar b Dc D\bar
c{\prod_{i}b(z_i)\bar b(z_i)\over \det B^{\circ n}(z_j,\bar
z_k)} \exp(-S[X] -{1\over2\pi}\int_C\sqrt g b\nabla^z_{1-n}c+c.c.)\
, \label{marvaillaise}\ee where $S[X]$ is the Polyakov action in
$2c_n$ dimensions, is a well-defined prescription on canonical curves. This implies that
\be \int_{\M_g}V[\J]=\int DgDXD\Psi \exp(-S[X]-S[\Psi]) \ ,
\label{formidabiledue}\ee where now the Polyakov action is in
$D=26+2\sum_{k\in \I}n_kc_k$
dimensions, is a Weyl anomaly free partition function. We have
\be V[\J]=\Z_{Pol}\prod_{k\in\I} V_{k}^{n_k}
\ , \label{volumiii}\ee
which are volume forms on $\M_g$. In particular, \cite{Matone:2012kk}
\be
V_{n}(\tau)={K_1^{(2n-1)^2}\over
K_n}{1\over(\det\tau_2)^{2n(n-1)}} \ ,
\label{eccoqua}\ee
which are a $(0,0)$-form on $\M_g$. Here
\be
{K}_n={\det B^{\circ n}(z_i,\bar z_j)\over |\det\phi_j^n(z_k)|^2}\bigl|\kappa[\phi^n]\bigr|^2 \ ,
\label{kenneaaa}\ee
where
the determinant of $B^{\circ n}(z_j,\bar z_k)$ is taken with the indices $j,k$ ranging from
$1$ to $N_n$. Note that
$$
K_1=
{|\kappa[\omega]|^2\over\det \tau_2} \ .
$$
One may check that $V_{n}(\tau)$ is obtained from the modulo square of the Mumford forms by a non-chiral mapping \cite{Matone:2012kk}.

We now look at the $\Z[\J]$ which are free of singularities at the Deligne-Mumford boundary. Following Fay
\cite{FayMAM}, we consider their behavior at $\partial\overline\M_g$.
Fay proved, using Bers like basis
$\phi^n_{t}=\{\phi^n_{i,t}\}_{i\in I_{N_n}}$ for $H^0(K_C^n)$, that
$F_{g,n}$ has a pole of order $n(n-1)/2$ in the plumbing fixture
parameter $t$.

Let us consider the degeneration limits. First, by approaching the generic
point on a
irreducible singular curve obtained by identifying two points $a,b$
on a smooth genus $g-1$ curve. In this case
\be
F_{g,n}[\phi^n_{t},\omega]\sim t^{-n(n-1)/2}{E(a,b)^{n-n^2}\over (2\pi
i)^{(2n-1)^2}}F_{g-1,n}[\phi^n,\omega] \ .
\label{everouno}\ee
In the case of degeneration corresponding to a reducible singular
curve obtained by identifying points on two smooth curves of genus
$g_1$ and $g-g_1$, we have
\be
F_{g,n}[\phi^n_{t},\omega]\sim \epsilon t^{-n(n-1)/2}
F_{g-g_1,n}[\phi^n\omega]G_{g_1,n}[\phi^n,\omega] \ ,
\label{everodue}\ee
where $\epsilon$ is a
fixed $(2g-2)$th root of unity.
We now consider rational functions of Mumford forms
\be
\F_g[\phi^k,\omega,\J]={\prod_{k\in \I}[F_{g,k}^{n_k}[\phi^k,\omega]\wedge^{max}\phi_j^k]\over (\omega_1\wedge\cdots\wedge\omega_g)^{d/2}} \ ,
\label{nercc}\ee
$n_k\in\ZZ$.
Note that
\be
d=2\sum_{k\in\I} n_k c_{k} \ .
\label{centraltotal}\ee
Since the Mumford forms on $\M_g$ are holomorphic and non-vanishing,
imposing the absence of singularities for $\F_g$ on $\overline\M_g$ is equivalent
to require holomorphicity at the Deligne-Mumford boundary. This gives
\be
\sum_{k\in\I} n_kk(k-1)={1\over6}\sum_{k\in\I} n_k(c_k-1)\leq0 \ .
\label{somma}\ee
Eq.(\ref{centraltotal}) and (\ref{somma}) imply
\be
d\leq2\sum_{k\in\I}  n_k \ .
\label{dsomma}\ee
We have

\begin{enumerate}

\item {\it
If $\sum_{k\in\I}n_kk(k-1)=0$,
then \be
\F_g[\phi^k,\omega,\J]=\F_g[\phi^k,\phi^1,\J] \ ,
\label{nerccbbb}\ee
with $\phi^1_1,\ldots,\phi_g^1$  an arbitrary basis of $H^0(K_C)$.
}

\item {\it $\sum_{k\in\I}n_kk(k-1)=0$ then at the non-separating node of the Deligne-Mumford boundary
\be
\F_g\sim {1\over (2\pi
i)^{d/2}}\F_{g-1} \ ,
\label{ewss}\ee
whereas at the separating node}
\be
\F_{g}\sim \epsilon
\F_{g-g_1}\F_{g_1} \ .
\label{everoduess}\ee

\end{enumerate}

\noindent
The invariance (\ref{nerccbbb}) follows by observing that if (\ref{somma}) is satisfied, then, due to the
term  $\det\phi^1_j(z_k)$ at the numerator of $\kappa[\phi^1]$
$$
\F_g=
\prod_{k\in\I}\Bigg({\wedge^{max}\phi_j^k\over
\kappa[\phi^k]}\Bigg)^{n_k} \ ,
$$
where $n_1\equiv -d/2$.
The factorization properties (\ref{ewss}) and (\ref{everoduess}) follow by (\ref{everouno}) and (\ref{everodue}).

\noindent
The previous analysis implies that the partition functions
$$
\Z[\J] \ ,
$$
with
\be
\sum_{k\in\I} n_k(c_k-1)+12\leq0 \ ,
\label{numbertheory}\ee
are finite at the Deligne-Mumford boundary. In this case
$$
D\leq2+2\sum_{k\in\I}n_k \ .
$$
The first partition function with $D=4$ space-time coordinates which is finite at $\partial\overline\M_g$ is the one with $n_2=0$, $n_3=n_4=1$ and $n_5=-1$, that is
$$
D=2(c_2+c_3+c_4-c_5)=4 \ ,
$$
corresponding to
\be
\int_{\M_g}{\cal Z}[1_3,1_4,-1_5]=\int DgDXD\Psi \exp(-S[X]-S[\Psi]) \ .
\label{formidabileddf}\ee
$S[X]$ is the four dimensional Polyakov action, whereas $S[\Psi]$ is the sum of the action of a non-chiral $b$-$c$ system of weight
$3$, another of weight $4$ and a $\beta$-$\gamma$ system of weight $5$.
Note that in this case
$$\sum_{k\in\I} n_k(c_k-1)+12=0 \ ,
$$ so that ${\cal Z}_g[1_3,1_4,-1_5]$ satisfies the factorization properties like the ones in (\ref{ewss}) and (\ref{everoduess}).

A similar analysis extends to $V[\J]$. However, since $\det B^{\circ n}(z_j,\bar z_k)$ vanishes on the hyperelliptic locus with $g\neq2$, the above results can be extended to $V[\J]$ in the case
of canonical curves. The extension to the full $\M_g$ requires some further condition when considering the hyperelliptic loci of $\M_g$. Let us stress that
$V[-1_2]$, that in \cite{Matone:2012kk} has been shown to coincide with the volume form on $\M_g$ induced by the Siegel metric,
corresponds to a finite string in $D=0$. It follows that
\be
Vol_S(\M_g)=\int_{\M_g} \Z[-1_2] \ ,
\label{volumeS}\ee
is the Siegel volume of $\M_g$.
This may indicate that, as discussed in \cite{Matone:2012wy} in the case of the Polyakov and supersymmetric strings, even the above models can be obtained, by a reduction
mechanism, from volume forms on the Siegel upper half-space.

The partition functions $\Z[\J]$ and $V[\J]$ represent an infinite class of volume forms on $\M_g$. Studying their symmetries and spectra may also lead to understand some of the
more recent questions of superstring perturbation theory \cite{D'Hoker:2001zp}-\cite{Witten:2012bh}.

\section*{Acknowledgements}

I thank  Giulio Bonelli, Giulio Codogni, Maurizio Cornalba, Paolo di Vecchia, Pietro Grassi, Samuel Grushevsky, Ian Morrison, Paolo Pasti, Riccardo Salvati Manni, Augusto Sagnotti, Dima Sorokin, Mario Tonin and Roberto Volpato for helpful comments and discussions.
This research is supported by the Padova
University Project CPDA119349 and by the MIUR-PRIN contract 2009-KHZKRX.

\end{document}